# Single-particle Fraunhofer diffraction in a classical pilot-wave model

David Darrow[*] and John W. M. Bush[†]

*Department of Mathematics, Massachusetts Institute of Technology, Cambridge, Massachusetts 02139, USA*

Walking oil droplets offer a qualitative, classical analog of single-particle diffraction. Making this analog quantitative has proven challenging, leading recent authors to conjecture that no classical pilot-wave model could exhibit Fraunhofer diffraction. We revisit the problem with the recent, Lagrangian pilot-wave model of Darrow and Bush [Symmetry **16**, 149 (2024)], and find agreement with both single- and double-slit Fraunhofer patterns. We identify two distinct dynamical features that enable our model to capture Fraunhofer diffraction and distinguish it from previous classical pilot-wave models.

## I. INTRODUCTION

Single- and double-slit diffraction has played a prominent role in the development of physics. Young's 1804 double-slit experiment [1] established a consensus around the wavelike nature of light, as formalized by Huygens [2] and Fresnel [3]. A century later, G. I. Taylor demonstrated slit diffraction using individual parcels of light [4], showing wavelike diffraction to be compatible with Einstein's *quantum* theory of light [5]. Inspired by the particle-wave duality of light, de Broglie proposed that matter should behave likewise [6], and his prediction of the wavelike diffraction of matter was soon confirmed by the experiments of Davisson and Germer [7].

It is natural to believe that single-particle diffraction is unique to quantum particles. For instance, Feynman claimed that the electron double-slit experiment is "absolutely impossible ... to explain in any classical way," and "has in it the heart of quantum mechanics" [8]. Such a viewpoint is motivated by two features of the experiment. First, interference arises even for single particles, so that diffraction cannot be explained by appealing to multiparticle interactions. Second, in a double-slit apparatus, a quantum particle is simultaneously affected by *both* slits, where the expectation is that a classical particle would only interact with one at a time.

The experiments of Couder and Fort offer a distinct, macroscopic example of both phenomena [10]. They performed single- and double-slit experiments with walking droplets, millimetric oil droplets self-propelling on the surface of a vibrating bath through interaction with their own wave field. Notably, this hydrodynamic system has commonalities with de Broglie's (now-historical) *double-solution* model

[11], in which a quantum particle was envisaged as having an internal vibration (at the Compton frequency) that excites a spatially extended guiding wave field [12].

Whichever slit a walking droplet passes through, its guiding wave is influenced by both slits, and so too is its trajectory. Couder and Fort reported a diffraction pattern comparable to the *Fraunhofer* pattern of quantum particles. However, subsequent work has shown that walking droplets are highly sensitive to experimental parameters not reported in Couder and Fort's original paper, including droplet size and the vertical acceleration of the bath; consequently, reproducing their diffraction results has proven elusive. Their results have also been contested on statistical grounds [13,14], and more refined, repeatable experiments [15,16] reveal quantitatively different diffraction patterns. These findings bring diffraction in line with the bulk of hydrodynamic quantum analogs [17], for which the correspondence is typically only qualitative.

The ensuing debate has raised the following question: While the diffraction patterns of walking droplets differ from those of quantum particles, might *any* classical pilot-wave model yield Fraunhofer patterns? Andersen *et al.* [13] and Bohr *et al.* [14] have conjectured that *no* classical pilot-wave system—that is, a model exhibiting a local, two-way coupling between particle and wave—could exhibit both single- and double-slit Fraunhofer diffraction. For such systems, they argue that particle-centered radiation would guarantee similar diffraction patterns from single- and double-slit apparatuses.

We here revisit the diffraction problem with our recent, Lagrangian pilot-wave model [9,18]. We show that this model yields quantitative agreement with both single- and double-slit Fraunhofer patterns. The width of the patterns is determined by an "effective" de Broglie wavelength $\lambda_{\text{eff}}$, proportional to $\lambda_{\text{dB}} = 2\pi\hbar/p$ for particle momentum $p$, so long as the velocity satisfies $u \lesssim 0.25c$. The constant of proportionality is prescribed by the model's only free parameter, the particle-wave coupling strength. Our results pass the statistical tests leveled by Andersen *et al.* [13] and Bohr *et al.* [14] against the results of Couder and Fort [10], and thus conclusively demonstrate that single-particle Fraunhofer diffraction is possible with a classical system.

[*]Contact author: ddarrow@mit.edu
[†]Contact author: bush@math.mit.edu

We enumerate the key physical mechanisms that allow our pilot-wave model to exhibit Fraunhofer diffraction. First, we examine the *Gedankenexperiment* that led Andersen *et al.* and Bohr *et al.* to their conjecture. We demonstrate that, while their argument applies to walking droplets and related pilot-wave models [19–21], the Lorentz-covariant radiation behavior of the present model circumvents it. Second, we demonstrate that, in order to recover Fraunhofer-like diffraction patterns in a classical pilot-wave system, particle trajectories must be either nonintersecting or chaotic. Our system obeys the latter criterion, while Bohmian mechanics—a nonlocal, pilot-wave interpretation of quantum mechanics [22] that also yields Fraunhofer diffraction patterns [23]—obeys the former. In addition to enabling Fraunhofer diffraction, Lorentz-covariant radiation and chaotic particle trajectories distinguish our pilot-wave model from its predecessors and open the door to a broader class of classical pilot-wave models.

## II. MATHEMATICAL MODEL

We first introduce our Lagrangian pilot-wave model [9] and review some of its key features. Our wave field is a real Klein-Gordon field $\phi$ with an associated mass $m$ and an applied energy potential $\widetilde{V}$; for convenience, we write $V = mc^2 + \widetilde{V}$. Our particle is a relativistic point particle at $\vec{q}_p \in \mathbb{R}^2$, also of mass $m$. The wave and particle are coupled through the following action, written in nondimensionalized units $m = \hbar = c = 1$:

$$\mathcal{S} = \frac{1}{2} \int d^2q\, dt\, (|\partial_t \phi|^2 - \|\nabla \phi\|^2 - V(q)^2 |\phi|^2)$$
$$- \int_0^{t'} dt\, \gamma^{-1}(1 + b^2/4\pi + b\phi(\vec{q}_p, t)),$$

where $\gamma = (1 - \|\vec{u}\|^2/c^2)^{-1/2}$ is the Lorentz factor of the particle, $\vec{u} = d\vec{q}_p/dt$ is its velocity, and $b > 0$ is the wave-particle coupling constant.

In our previous work [9], we showed that, for sufficiently small coupling ($b \lesssim 25.0$), the Euler-Lagrange equations take the following form [24]:

$$\left(\partial_t^2 - \nabla^2 + V(\vec{q})^2\right)\phi = \gamma^{-1} b \delta^2(\vec{q} - \vec{q}_p),$$
$$d_t(\gamma \vec{u}) = \gamma^{-1} b \nabla \phi(\vec{q}_p, t). \quad (1)$$

Nondimensionalizing reveals that the only free parameter in our system is the coupling constant $b$. Notably, if $b = 0$, the particle and wave decouple completely.

The units we have selected are those of Compton; they select the Compton wavelength $\lambda_c = 2\pi[\hbar/mc]$ as the natural length scale and the Compton period $T_c = 2\pi[\hbar/mc^2] = 2\pi/\omega_c$ as the natural timescale. As in de Broglie's mechanics [11], this should be seen as a "fast scale" for our system, over which wave radiation and the equilibration of particle-wave energy occur. The emergent length scale of the pilot wave field is the *de Broglie* wavelength $\lambda_{dB} = \lambda_c/\gamma u$ (see Ref. [9]), which determines the length scale of the system's diffraction patterns.

A rigorous derivation of Eq. (1) is presented in our previous work (Appendix B of Ref. [9]). In short, this derivation rests on the assumption of a small coupling constant, allowing us to neglect a wave-induced correction to the particle's inertial mass. We showed numerically that this correction is negligible when $b \lesssim 25.0$, so we focus on the same parameter regime in the present work.

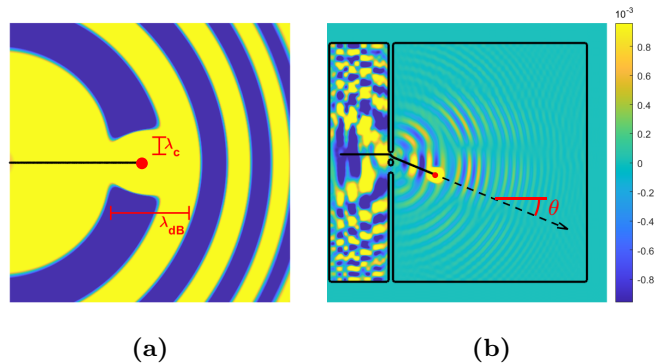

FIG. 1. (a) A close-up of the free particle in a steady state, reproduced from Ref. [9]. The image is color-coded according to the signed amplitude $\phi$ of the pilot wave. The particle radiates energy only when it accelerates, but is always accompanied by a Compton-scale wave packet and a quasimonochromatic pilot wave field with the de Broglie wavelength. (b) A trajectory in our double-slit experiment shows the particle exiting the slit at a diffraction angle $\theta$. The standing wave pattern apparent behind the slit only materializes once the particle has exited the slit, so does not influence its dynamics.

In the case of a free particle, the system [Eq. (1)] recovers several of the key dynamical features hypothesized in de Broglie's double-solution [9], as highlighted in Fig. 1. First, it exhibits emergent particle vibrations at the redshifted Compton frequency $\gamma^{-1}\omega_c$, consistent with the *Zitterbewegung* imagined by de Broglie [25]. It also satisfies the de Broglie relation $\vec{p} = \hbar \vec{k}$ at the particle's position; heuristically, this can be seen by equating the particle's velocity and the group velocity of its (locally monochromatic) pilot wave. Taken in conjunction, these two properties imply de Broglie's *harmony of phases*, in which the vibrations of the particle and wave are synchronized in any frame of reference. Finally, the particle is always tracked by a robust Compton-scale wave packet, independent of the particle's velocity or acceleration—apart from relativistic length contraction by a factor $\gamma^{-1}$ in the particle's direction of motion.

A key feature of this dynamical system is its radiation behavior: The particle radiates energy only upon *acceleration*, similar to the case of classical electromagnetism [Fig. 1(a)]. This behavior is necessitated by the Lorentz covariance of our pilot wave. We demonstrate below how this feature distinguishes our system from prior pilot-wave models and allows for Fraunhofer diffraction [26].

## III. DIFFRACTION RESULTS

We now turn to our diffraction simulations. Our double-slit setup is shown in Fig. 1(b). The particle starts on the left-hand side of the domain, a distance $13.8\lambda_c$ to the left of the slits with an initial momentum $p_0$ and impact parameter $y$ (see Fig. 2). The initial momentum is quickly (on the Compton scale $T_c$) equilibrated between the particle and wave, leaving the particle with a unique steady-state velocity $\vec{u} = \vec{u}(p_0)$ and accompanying de Broglie waves [see Fig. 1(a)]. The details of the momentum exchange and ensuing radiation are explored at length in our previous work [9], but are not critical here. Eventually, the particle passes through one of the slits and exits on a different rectilinear path. We fix $V(\vec{q})^2 = 1mc^2$ outside the boundaries of the apparatus, corresponding to zero external potential. The walls of the apparatus are $0.65\lambda_c$ thick, with $V(\vec{q})^2 = 5m^2c^4$, ensuring that waves cannot penetrate them. The wall potential $V$ is smoothed by a kernel of width $\lambda_c/2$.

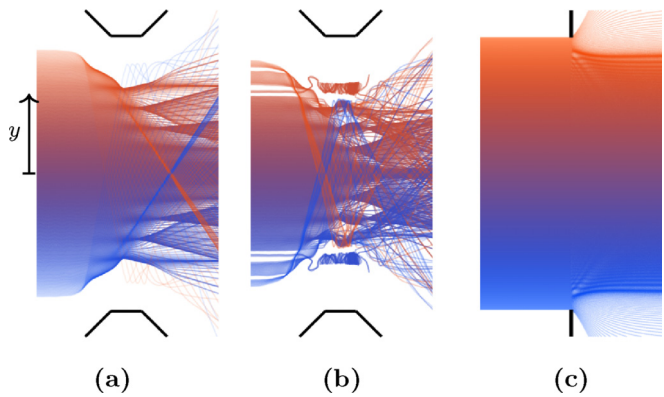

FIG. 2. Single-slit trajectories with slit width $w = 4.07\lambda_c$, initial momentum $p_0 = 0.3m\gamma$, 500 evenly spaced impact parameters $|y| < w/2$, and coupling constants (a) $b = 16.7$ and (b) $b = 25.0$. Trajectories cross in regular fold patterns, with spacing $0.28\lambda_c$ between folds in both simulations. (c) Bohmian trajectories deduced for matching slit width and wavelength. Notably, Bohmian trajectories do not cross one another in diffraction experiments [22,23].

In Figs. 2(a) and 2(b), we show trajectories for two single-slit experiments, with coupling strengths $b = 16.7$ and $b = 25.0$. Our trajectories differ from those of Bohmian mechanics [Fig. 2(c)], which were calculated first by Philipiddis [23]. For one, they cross in regular fold patterns, in a manner reminiscent of walking droplet diffraction experiments [16,27,28]. Further investigation of these trajectories is provided in Appendix A. The spacing between folds ($\sim 0.28\lambda_c$) is independent of $b$, but as $b$ increases, chaotic behavior emerges at the Compton scale $\lambda_c$.

In a single- or double-slit experiment, if $w$ is the width of a single slit and $d$ is the separation between two slits, an incoming plane wave of wavelength $\lambda$ diffracts according to a Fraunhofer probability density

$$\rho(\theta) \propto \cos^2(\pi d \sin(\theta)/\lambda) \operatorname{sinc}^2(\pi w \sin(\theta)/\lambda), \quad (2)$$

where $\theta$ is the far-field diffraction angle, and $d = 0$ for the single-slit case. We will proceed by demonstrating that our system obeys Eq. (2), provided we choose the following *effective* de Broglie wavelength $\lambda_{\text{eff}} = \lambda_{\text{eff}}(b)$, prescribed by the particle momentum $p$ and coupling constant $b$ in a manner characterized in Appendix B:

$$\lambda_{\text{eff}}(b) \sim (b/68.0)^2 \lambda_{\text{dB}} = (b/68.0)^2 (2\pi\hbar/p). \quad (3)$$

In Figs. 3(a)–3(c), we show our single-slit diffraction results for a range of $b$. Figures 3(a) and 3(c) correspond to the trajectories reported in Figs. 2(a) and 2(b). We weight the distribution of impact parameters by a Gaussian of standard deviation $0.41\lambda_c$, centered on the slit. Alongside, we show the corresponding Fraunhofer patterns [Eq. (2)] with wavelength [Eq. (3)]. To simulate classical uncertainty in the initial particle state, we convolve the prediction [Eq. (2)] by a Gaussian of standard deviation

$$\sigma := 0.02\lambda_{\text{eff}} w/\lambda_c^2 \quad (\text{rad}), \quad (4)$$

fitted to single-slit data with $b = 16.7$ and $p_0 = 0.3m\gamma$. We note that such smoothing is typical in the analysis of laboratory diffraction experiments [29,30].

In Fig. 3(d), we show the diffraction pattern arising from our double-slit experiment. Of the 2500 total simulations, only $N_{\text{good}} = 1296$ particles cross the apparatus without being reflected back. Trajectories are again weighted by a Gaussian of standard deviation $0.41\lambda_c$ at the center of each slit. Alongside are two possible fits to the data: the Fraunhofer prediction [Eq. (2)] with the smoothing [Eq. (4)] and a Gaussian of the correct variance. We compare the $\chi^2$ values of each, using both the Pearson estimate $\chi_P^2$ and the Yates-corrected estimate $\chi_Y^2$ [31]. If $\nu = n_b - n_c$ is the number of degrees of freedom—the bin

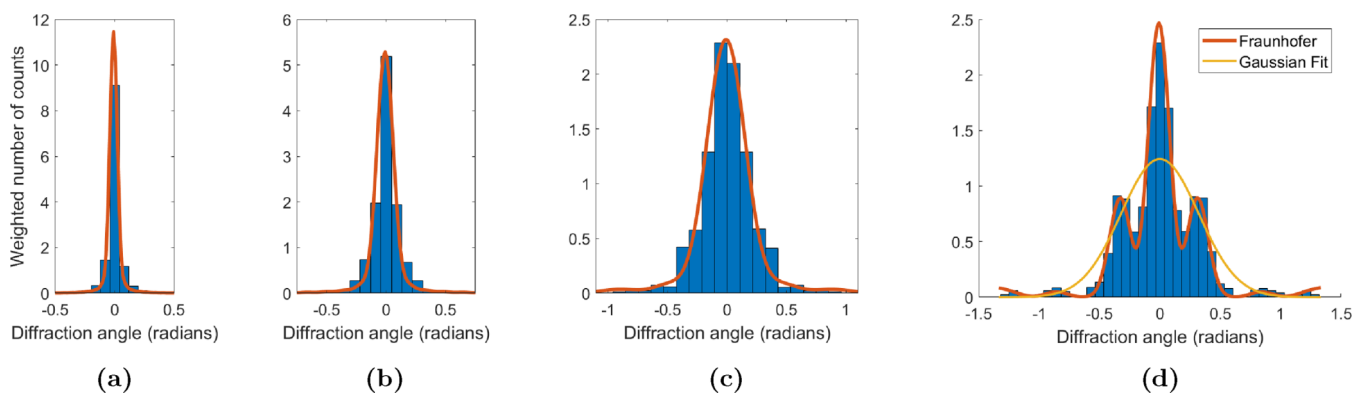

FIG. 3. Single-slit diffraction patterns arising in our pilot-wave system, with initial momentum $p_0 = 0.3m\gamma$, slit width $w_1 = 4.07\lambda_c$, and coupling parameters (a) $b = 16.7$, (b) 20.9, and (c) 25.0. Particle trajectories corresponding to panels (a) and (c) are reported in Figs. 2(a) and 2(b). The Fraunhofer predictions [Eq. (2)] are shown for comparison, smoothed as indicated in Eq. (4). (d) A double-slit diffraction pattern with $b = 25.0$, made up of 2500 runs with initial momentum $p_0 = 0.3m\gamma$, slit width $w_2 = 2.03\lambda_c = w_1/2$, and slit separation $d = 3.66\lambda_c$, compared to a Fraunhofer curve of wavelength $\lambda_{\text{eff}}$ and smoothing [Eq. (4)], and to a Gaussian of the same variance.

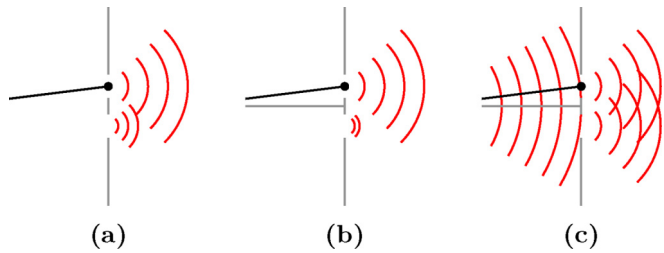

FIG. 4. The *Gedankenexperiment* of Andersen *et al.* [13] and Bohr *et al.* [14]. (a) They argue that, for classical pilot-wave systems, continuous particle radiation imbalances the wave energy incident on the two slits. (b) One expects this imbalance to be magnified by introducing a wall in front of the apparatus: For a sufficiently long wall, the double-slit diffraction pattern should converge to that of a single-slit. (c) Since our particle emits radiation primarily at its point of origin, far from the apparatus, the incident wave energy is comparable at both slits, even in the presence of a wall. Our system thus circumvents their conjecture (Conjecture 1).

count less the number of constraints—then $\chi^2/\nu \sim 1$ for a good fit [32]. We choose $n_b = 37 \sim \sqrt{N_{\text{good}}}$, and note that $n_c = 1$ for both our Fraunhofer curve (i.e., the smoothness $\sigma$) and the Gaussian (its variance). Both $\chi^2$ values confirm a close adherence to the Fraunhofer prediction:

|  | Fraunhofer | Gaussian |
| --- | --- | --- |
| $\chi_P^2/\nu$ | 1.36 | 17.6 |
| $\chi_Y^2/\nu$ | 0.89 | 11.0 |

Numerical limitations prevent us from further reducing the ratio $w/\lambda_{\text{eff}}$, which would strengthen the secondary Fraunhofer nodes. Specifically, with smaller $w/\lambda_{\text{dB}}$, more particles reflect backward off the slit, so it becomes increasingly difficult to achieve statistically robust particle counts.

## IV. THE LIMITS OF PILOT-WAVE DIFFRACTION

We proceed by identifying two physical mechanisms—both necessary for Fraunhofer diffraction—that distinguish the present pilot-wave model from its predecessors. First, we attempt to formalize the conjecture of Andersen *et al.* [13] and Bohr *et al.* [14] as follows.

*Conjecture 1.* If a pilot-wave system is local, and the particle is the only source of the wave, then it cannot exhibit both single- and double-slit Fraunhofer diffraction.

They argue this point with the following *Gedankenexperiment*, which we depict in Fig. 4. They argue that, in any classical pilot-wave system, the particle's continuous radiation creates an energy imbalance in the double-slit apparatus, with stronger waves propagating through the same slit as the particle [Fig. 4(a)]. This imbalance can be magnified by adding a wall between the two slits [Fig. 4(b)]; the field is then excited on the particle's side of the wall and dispersed on the other. The longer the wall, the weaker the effect of the more distant slit on the particle, and the more closely the diffraction pattern conforms to that of a single slit. In the presence of such a wall, classical pilot-wave systems would yield similar diffraction patterns in single- and double-slit arrangements.

Their argument holds for walking droplets (and related classical pilot-wave models [19–21]), in which waves are excited continuously along the particle path and dissipated elsewhere. While coherent double-slit diffraction patterns have been observed with walking droplets [15,16], this system does have the form hypothesized by Anderson *et al.* and Bohr *et al.*, so should not be expected to yield Fraunhofer diffraction in both single- and double-slit experiments.

In contrast, the particle in our pilot-wave model radiates only while *accelerating* [9]. It thus excites a de Broglie wave front at its original point of acceleration, far from the apparatus, and does not excite more waves until it accelerates again in passing through the slit. As such, symmetry is approximately maintained between the wave energy incident on the two slits [Fig. 4(c)]. This symmetry is apparent in the double-slit geometry depicted in Fig. 1(b). While the modified geometry of Anderson *et al.* and Bohr *et al.* [i.e., the insertion of a wall, as in Fig. 4(b)] would require a far larger simulation domain than we can practically utilize, one expects that the same should hold in that case. In short, Conjecture 1 does not apply to our model.

In Appendix A, we prove another requirement for smooth diffraction patterns (such as that of Fraunhofer) in a pilot-wave system:

> To exhibit diffraction patterns without sharp peaks, particle trajectories must either (1) not cross one another or (2) exhibit chaos.

The first criterion is satisfied by Bohmian mechanics and the second is satisfied by our pilot-wave system, as we show in Appendix A.

## V. DISCUSSION

In this work, we have demonstrated that our Lagrangian pilot-wave model [9] replicates the single- and double-slit Fraunhofer diffraction patterns of quantum mechanics. Our particle carries an *effective* de Broglie wavelength $\lambda_{\text{eff}} \propto \lambda_{\text{dB}}$, with proportionality constant prescribed [as in Eq. (3)] by the model's only free parameter, the particle-wave coupling constant $b$.

Our results rebut Conjecture 1 of Andersen *et al.* [13] and Bohr *et al.* [14], namely, that no classical pilot-wave system can exhibit Fraunhofer diffraction. Although their conjecture does preclude Fraunhofer patterns with walking droplets and related models [19–21], we have shown how the Lorentz-covariant radiation behavior of our model circumvents their argument. In its place, we derived a different, necessary criterion for smooth, Fraunhofer-like diffraction patterns, which highlights the critical role of *chaos* in our results. These features distinguish the present model from previous pilot-wave systems, showing that classical pilot-wave dynamics is far richer than previously envisioned.

Looking forward, we anticipate that our model will be helpful in capturing a variety of quantumlike behaviors out of reach of the walking droplet system. For example, Fort and Couder [33] studied a system of "inertial walkers," wherein the radiation emitted by the walker is stationary with respect to its rest frame (rather than the laboratory frame), and recovered an analog of the Bohr-Sommerfeld quantization rule for orbiting inertial walkers. The present system naturally recovers exactly this form of radiation, and we anticipate that similar orbital quantization conditions might emerge from it. On the other hand, since our model (slowly) radiates energy upon particle acceleration, we do not expect its long-term statistics to converge to "quantumlike" results, even though such a phenomenon is ubiquitous in the droplet system [17,34]. Moreover, without nonlocality (a critical feature of Bohmian mechanics [22]), the work of Bell [35] and its subsequent experimental verification [36] imply that there is no way for any classical pilot-wave model to capture multiparticle entanglement. Nevertheless, in demonstrating single-particle Fraunhofer diffraction in a classical system, we have shifted the boundary between quantum and classical physics, and opened the door to a class of local, classical pilot-wave models.

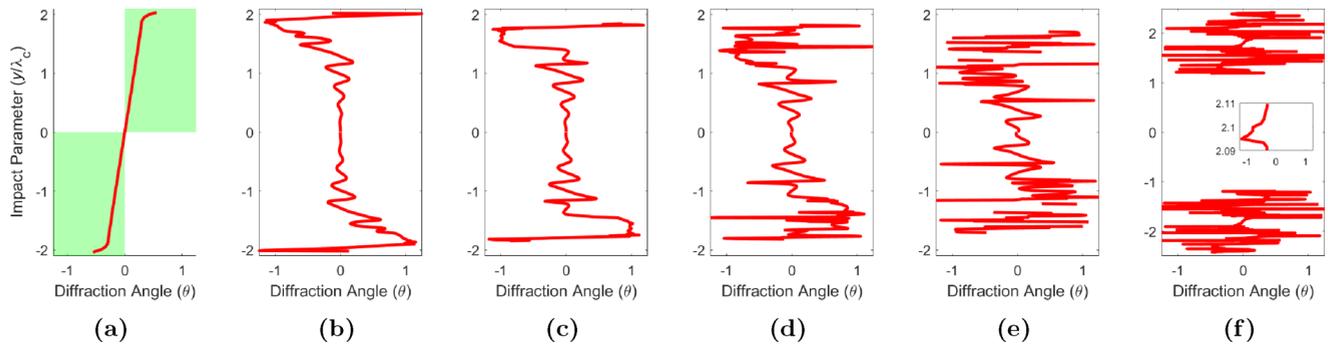

FIG. 5. Diffraction maps relate incoming impact parameter $y_{in}$ to far-field diffraction angle $\theta_{out}$, measured in radians. (a) Bohmian diffraction map corresponding to Fig. 2(c). All trajectories are "surreal," and so do not cross the system's centerline; Bohmian trajectories thus satisfy the first criterion of Proposition A1, allowing them to exhibit Fraunhofer diffraction. (b)–(e) Diffraction maps corresponding to our single-slit simulations, with slit width $w_1 = 4.07\lambda_c$, initial momentum $p_0 = 0.3m\gamma$, and coupling strengths $b = 12.5, 16.7, 20.9, 25.0$. As is evident in Figs. 2(a) and 2(b), as $b$ increases, regular fold patterns are replaced by chaotic behavior; our system thus satisfies the second criterion of Proposition A1. (f) A diffraction map corresponding to our double-slit experiment in Fig. 3(d), with slit width $w_2 = 2.03\lambda_c = w_1/2$, slit separation $d = 3.66\lambda_c$, initial momentum $0.3m\gamma$, and coupling strength $b = 25.0$. A cutout at $2.09 < y < 2.11$ reveals that the map is continuous but not smooth. Since it is not smooth, our model is able to exhibit a smooth (e.g., Fraunhofer) diffraction pattern.

## ACKNOWLEDGMENTS

The authors gratefully acknowledge the Office of Naval Research for financial support via Grant No. N00014-24-1-2232. D.D. acknowledges support from an NDSEG Graduate Fellowship.

## DATA AVAILABILITY

The data that support the findings of this article are not publicly available upon publication because it is not technically feasible and/or the cost of preparing, depositing, and hosting the data would be prohibitive within the terms of this research project. The data are available from the authors upon reasonable request.

## APPENDIX A: PARTICLE TRAJECTORIES

To better understand what distinguishes our model from existing pilot-wave systems, it is instructive to study the particle trajectories that make up our diffraction results. We introduce the *diffraction map*,

$$S : y_{\text{in}} \mapsto \theta_{\text{out}}, \tag{A1}$$

that sends impact parameters $y_{\text{in}}$ to diffraction angles $\theta_{\text{out}}$. This map analogizes the quantum *scattering matrix*, and provides a touchstone between distinct pilot-wave models.

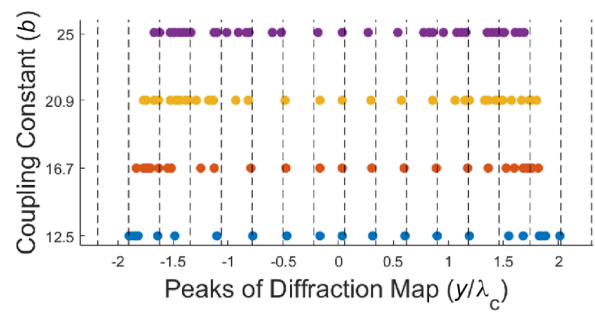

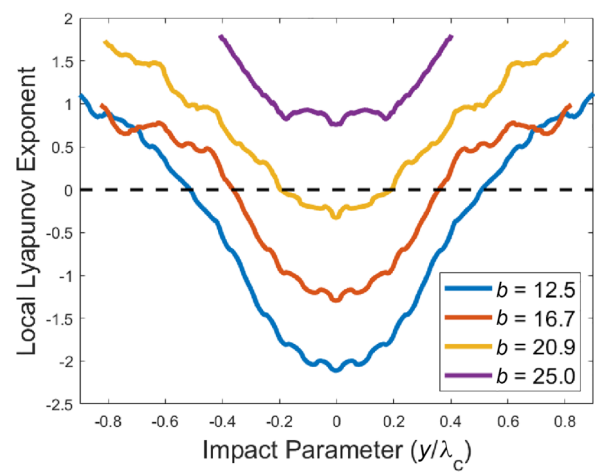

FIG. 6. (a) Peaks of the single-slit diffraction maps of Figs. 5(a)–5(d), along with an equispaced grid fit to the central peaks of the $b = 12.5$ map (i.e., those with $y/\lambda \in [-1.2, 1.2]$). The spacing between peaks (approximately $0.28\lambda_c$) appears independent of the coupling parameter, except in the chaotic regions near the slit edges. (b) Local Lyapunov exponent for the four single-slit diffraction maps. As $b$ increases, chaos emerges first at the edges of the slit, then eventually spans the entire domain.

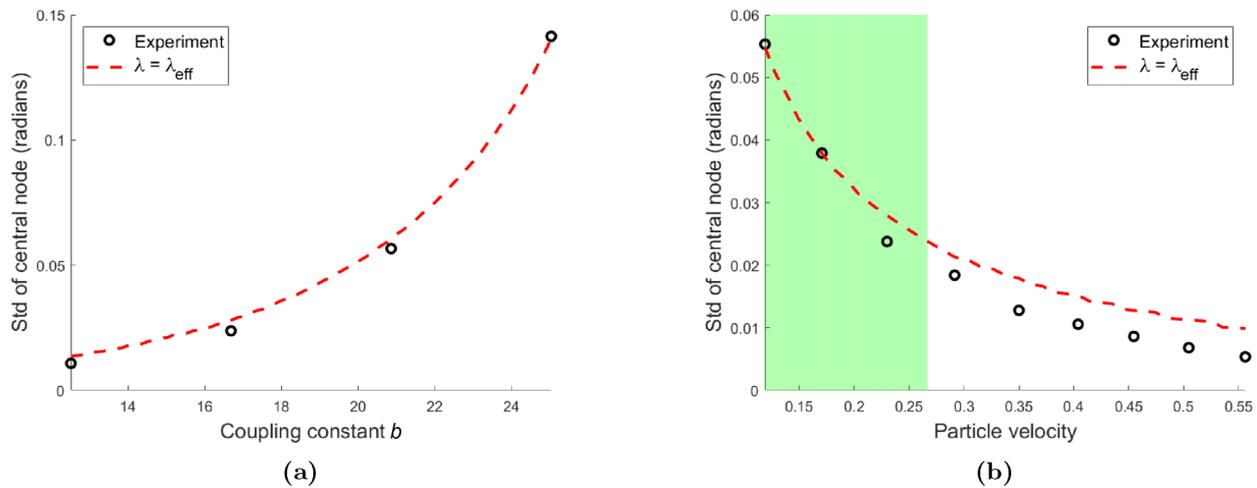

FIG. 7. (a) Dependence of the width of the central single-slit Fraunhofer node (black circles) on the coupling strength $b$ for fixed initial momentum $p_0 = 0.3m\gamma$, slit width $w = 4.07\lambda_c$, and 500 evenly spaced impact parameters $|y| < w/2$, compared to Fraunhofer curves (red) of wavelength [Eq. (3)] and smoothing [Eq. (4)]. The strong alignment indicates that our estimate of the effective de Broglie wavelength (which implies $\lambda_{\text{eff}} \propto b^2$) holds closely in this parameter regime. (b) Dependence of the central node width on steady-state velocity, for fixed $b = 16.7$. The estimate [Eq. (3)] (which implies $\lambda_{\text{eff}} \propto \lambda_{\text{dB}}$) holds only for slow-moving particles, highlighted in green; the effective de Broglie wavelength $\lambda_{\text{eff}}$ falls off faster than $\lambda_{\text{dB}}$ for fast-moving particles.

In Figs. 5(b)–5(e), we show single-slit diffraction maps for a range of coupling parameters $b$. As evident in Fig. 2, these trajectories are characterized by regular fold patterns at low values of $b$, but give way to chaotic behavior as $b$ increases. We show the peaks of these diffraction maps in Fig. 6(a), along with an equispaced grid (in $y$) fit to the central peaks of the $b = 12.5$ map (i.e., those with $y/\lambda \in [-1.2, 1.2]$). The spacing between peaks (approximately $0.28\lambda_c$) appears independent of the coupling parameter—except in the chaotic regions near the edge.

We investigate the emergence of chaos by looking at the *Lyapunov exponent* of our diffraction maps [37]:

$$\ell = \frac{1}{w} \int dy \, \ln \left| \lambda_c \frac{dS}{dy} \right|. \tag{A2}$$

The Lyapunov exponent quantifies the rate at which nearby trajectories diverge. If $\ell > 0$, it indicates that the system is chaotic, while if $\ell < 0$ it indicates that the system is diffusive. We show the *local* Lyapunov exponent of our four single-slit diffraction maps in Fig. 6(b). Here, the average over all impact parameters in Eq. (A2) is replaced by a local average over a window of width $\lambda_c$. All four experiments exhibit chaotic trajectories near the edges of the slit. Chaotic behavior becomes more prevalent as $b$ increases, and is present across the full domain in the $b = 25.0$ case.

Finally, the diffraction map gives a necessary criterion for pilot-wave systems to exhibit diffraction patterns without sharp peaks (such as that of Fraunhofer).

*Proposition A1.* Suppose a family of particle trajectories is parametrized by a one-dimensional parameter $y \in I$, where $I \subset \mathbb{R}$ is an open set, and that $y$ is distributed according to a nonsingular probability density $\rho_{\text{in}}$ on $I$. Consider the diffraction map $S : y \mapsto \theta$ that maps each configuration to its deterministic diffraction angle. To exhibit a smooth and bounded probability density $\rho_{\text{out}}(\theta)$, such as that of Fraunhofer, one of the following conditions must be met:

(1) Whenever $\rho_{\text{in}}(y) > 0$, we have $S'(y) \neq 0$.
(2) The map $S$ is nondifferentiable.

In short, any smooth extrema in the diffraction map correspond to sharp peaks in the diffraction pattern, since $|\frac{d\theta}{dy}| = |S'(y)|^{-1} = +\infty$; Proposition A1 gives sufficient conditions to avoid this occurrence.

The first of these criteria is satisfied by Bohmian trajectories; since they cannot cross one another, they generically satisfy $S'(y) > 0$ everywhere, as is apparent in Fig. 5(e). The second criterion is satisfied by the present model, through the emergence of chaos. In contrast, the walking droplet experiments of Ellegaard and Levinsen [16] do not appear to satisfy either criterion; particle trajectories cross one another (so $dS/dy = 0$ at points), but are not generally chaotic. We note that the earlier work of Pucci *et al.* [15] did report occasionally chaotic droplet behavior at near-critical system vibration. Nevertheless, both studies reported diffraction patterns with relatively sharp peaks.

We provide a proof of Proposition A1 below. Note that, while we discuss the proposition in terms of Fraunhofer diffraction, it provides a necessary criterion for a particle-based model to recover *any* smooth diffraction pattern.

*Proof of Proposition A1.* Note that $\rho_\text{out}$ is given by a pushforward measure:

$$\rho_\text{out}(\theta) = S_*\rho_\text{in}(y) = \rho_\text{in}(S^{-1}(\theta)).$$

Now, suppose $S$ is everywhere differentiable, and, for some $y_0 \in I$, we have both $\rho_\text{in}(y_0) > 0$ and $S'(y_0) = 0$. Let $B_\varepsilon^y = [y_0 - \varepsilon, y_0 + \varepsilon]$ for $\varepsilon > 0$; since $\rho_\text{in}(y_0) \neq 0$, we know that $\rho_\text{in}(B_\varepsilon) > C\varepsilon$ for sufficiently small $\varepsilon$, with $C > 0$. As $S'(y_0) = 0$, Taylor's theorem tells us that

$$|S(y) - S(y_0)| < \varepsilon f(\varepsilon)$$

for all $y \in B_\varepsilon^y$, where $f(\varepsilon) \to 0$ as $\varepsilon \to 0$. Fix a sequence $\varepsilon_i \to 0$, set $\delta_i = \varepsilon_i f(\varepsilon_i)$, and define $B_\delta^\theta = [S(y_0) - \delta, S(y_0) + \delta]$ for every $\delta > 0$. Then, we know that $S(B_{\varepsilon_i}^y) \subset B_{\delta_i}^\theta$ for all $i$, and thus

$$\frac{\rho_\text{out}(B_{\delta_i}^\theta)}{\delta_i} > \frac{\rho_\text{in}(B_{\varepsilon_i}^y)}{\delta_i} > \frac{c\varepsilon_i}{\delta_i}.$$

But $\varepsilon_i/\delta_i \to \infty$ as $\varepsilon_i \to 0$, so by the Radon-Nikodym theorem, the measure $\rho_\text{out}$ is singular. ∎

## APPENDIX B: THE EFFECTIVE DE BROGLIE WAVELENGTH

We proceed by rationalizing the form of the diffraction patterns exhibited by our model, by examining how the diffraction behavior depends on system parameters.

By adapting Noether's theorem to the case of pilot waves [9], we have shown previously that the horizontal momentum imparted from the field to the particle scales as $b^2$. As such, we expect the horizontal extent of our diffraction patterns—here quantified by the wavelength $\lambda_\text{eff}$—to scale likewise. We further hypothesize that $\lambda_\text{eff}$ scales with the wavelength of the particle's local wave field, which is exactly the de Broglie wavelength $\lambda_\text{dB} = 2\pi\hbar/m\gamma u$ [9]. In short, we predict $\lambda_\text{eff} \propto b^2 \lambda_\text{eff}$, so it remains only to calculate the constant of proportionality.

In Fig. 7, we show that the choice [Eq. (3)] is appropriate over a large parameter regime. In Fig. 7(a), we perform single-slit diffraction experiments over a range of $b$; three of these experiments correspond to the data reported in Figs. 3(a)–3(d). We report the standard deviation of *the central node* of each diffraction pattern, i.e., the pattern restricted to $|\theta| \leqslant \sin^{-1}(\lambda_\text{eff}/w)$. This metric is more robust than the total standard deviation, as it cuts out outliers and secondary nodes when they appear. We compare these results to those of the corresponding Fraunhofer predictions, with wavelength [Eq. (3)] and smoothing [Eq. (4)]. We see close correspondence to the $b^2$ scaling law, with the scaling coefficient as prescribed by Eq. (3).

In Fig. 7(b), we now fix $b = 16.7$ and carry out single-slit diffraction experiments over a range of velocities $u$. We see now that the scaling $\lambda_\text{eff} \propto \lambda_\text{dB}$ is a good approximation only for slow-moving particles (depicted in green), and that $\lambda_\text{eff}$ falls off faster than $\lambda_\text{dB}$ as $u$ increases. All simulations in the main text are in the parameter regime of Fig. 7(a), where the estimate [Eq. (3)] is satisfied.

[1] R. S. G. Britain, *Philosophical Transactions of the Royal Society of London* (W. Bowyer and J. Nichols for Lockyer Davis, printer to the Royal Society, London, United Kingdom, 1804), Vol. 94.

[2] C. Huygens, *Traité de la Lumière*, *Les Maitres de la Pensée Scientifique* (Gauthier-Villars, Paris, France, 1920).

[3] A. Fresnel, Memoire sur la diffraction de la lumiere in *Mémoires de l'Académie des sciences, tome V* (Crochard, Paris, 1819), pp. 339–475.

[4] G. I. Taylor, Interference fringes with feeble light, Proc. Camb. Philos. Soc. **15**, 114 (1909).

[5] A. Einstein, Über einen die Erzeugung und Verwandlung des Lichtes betreffenden heuristischen Gesichtspunkt, Ann. Phys. **322**, 132 (1905).

[6] L. De Broglie, Recherches sur la théorie des quanta, Ann. Phys. **10**, 22 (1925).

[7] C. J. Davisson and L. H. Germer, Reflection of electrons by a crystal of nickel, Proc. Natl. Acad. Sci. USA **14**, 317 (1928).

[8] R. P. Feynman, R. B. Leighton, and M. Sands, *The Feynman Lectures on Physics: The Definitive Edition* (Pearson, San Francisco, 2009), Vol. 3.

[9] D. Darrow and J. W. M. Bush, Revisiting de Broglie's double-solution pilot-wave theory with a Lorentz-covariant Lagrangian framework, Symmetry **16**, 149 (2024).

[10] Y. Couder and E. Fort, Single-particle diffraction and interference at a macroscopic scale, Phys. Rev. Lett. **97**, 154101 (2006).

[11] L. de Broglie, Interpretation of quantum mechanics by the double solution theory, Ann. Fond. Louis de Broglie **12**, 1 (1987).

[12] J. W. M. Bush, Pilot-wave hydrodynamics, Annu. Rev. Fluid Mech. **47**, 269 (2015).

[13] A. Andersen, J. Madsen, C. Reichelt, S. Rosenlund Ahl, B. Lautrup, C. Ellegaard, M. T. Levinsen, and T. Bohr, Double-slit experiment with single wave-driven particles and its relation to quantum mechanics, Phys. Rev. E **92**, 013006 (2015).

[14] T. Bohr, A. Andersen, and B. Lautrup, Recent Advances in Fluid Dynamics with Environmental Applications in *Bouncing Droplets, Pilot-Waves, and Quantum Mechanics*, edited by J. Klapp, L. Di G. Sigalotti, A. Medina, A. López, and G. Ruiz-Chavarra, (Springer, Cham, Switzerland, 2016), pp. 335–349.

[15] G. Pucci, D. M. Harris, L. M. Faria, and J. W. M. Bush, Walking droplets interacting with single and double slits, J. Fluid Mech. **835**, 1136 (2018).

[16] C. Ellegaard and M. T. Levinsen, Interaction of wave-driven particles with slit structures, Phys. Rev. E **102**, 023115 (2020).

[17] J. W. M. Bush and A. U. Oza, Hydrodynamic quantum analogs, Rep. Prog. Phys. **84**, 017001 (2021).

[18] D. Darrow, Convergence to Bohmian mechanics in a de Broglie-like pilot-wave system, Found. Phys. **55**, 13 (2025).

[19] M. Durey and J. W. M. Bush, Classical pilot-wave dynamics: The free particle, Chaos **31**, 033136 (2021).

[20] Y. Dagan and J. Bush, Hydrodynamic quantum field theory: The free particle, C. R. Mec. **348**, 555 (2020).

[21] M. Durey and J. Bush, Hydrodynamic quantum field theory: The onset of particle motion and the form of the pilot wave, Front. Phys. **8**, 300 (2020).

[22] P. R. Holland, *The Quantum Theory of Motion: An Account of the de Broglie-Bohm Causal Interpretation of Quantum Mechanics* (Cambridge University Press, Cambridge, UK, 1993).

[23] C. Philippidis, C. Dewdney, and B. J. Hiley, Quantum interference and the quantum potential, Nuov. Cim. **52**, 15 (1979).

[24] We have changed units slightly from our previous work [9], for clarity; the range $b \in (12.5, 25.0)$ we study here corresponds to the previously reported values $b_{\text{old}} \in (40.0, 80.0)$.

[25] D. Hestenes, Zitterbewegung structure in electrons and photons, arXiv:1910.11085.

[26] We note that an alternative choice of particle-wave coupling was investigated by Darrow [18], wherein the particle is coupled to the complex phase of $\phi$. He showed that, in the nonrelativistic limit $u \ll c$, this system converges exactly to single-particle Bohmian mechanics. Like Bohmian mechanics, then, this limit must exhibit Fraunhofer diffraction [23]. However, since wave propagation speed grows to infinity in this limit, it does not provide a compelling example of local, classical pilot-wave diffraction, so we consider it no further.

[27] B. K. Primkulov, D. J. Evans, V. Frumkin, P. J. Sáenz, and J. W. M. Bush, Diffraction of walking drops by a standing Faraday wave, Phys. Rev. Res. **7**, 013226 (2025).

[28] G. Pucci, A. Bellaigue, A. Cirimele, G. Alì, and A. U. Oza, Single-particle diffraction with a hydrodynamic pilot-wave model, Phys. Rev. E **111**, L033101 (2025).

[29] A. Tonomura, J. Endo, T. Matsuda, T. Kawasaki, and H. Ezawa, Demonstration of single-electron buildup of an interference pattern, Am. J. Phys. **57**, 117 (1989).

[30] A. H. Tavabi, C. B. Boothroyd, E. Yücelen, S. Frabboni, G. C. Gazzadi, R. E. Dunin-Borkowski, and G. Pozzi, The Young-Feynman controlled double-slit electron interference experiment, Sci. Rep. **9**, 10458 (2019).

[31] F. Yates, Contingency tables involving small numbers and the $\chi^2$ test, Suppl. J. R. Stat. Soc. **1**, 217 (1934).

[32] P. R. Bevington and D. K. Robinson, *Data Reduction and Error Analysis for the Physical Sciences*, 3rd ed. (McGraw-Hill, New York, 2003).

[33] E. Fort and Y. Couder, Trajectory eigenmodes of an orbiting wave source, Europhys. Lett. **102**, 16005 (2013).

[34] M. Durey, P. A. Milewski, and Z. Wang, Faraday pilot-wave dynamics in a circular corral, J. Fluid Mech. **891**, A3 (2020).

[35] J. S. Bell, On the Einstein–Podolsky–Rosen paradox, Phys. Phys. Fiz. **1**, 195 (1964).

[36] A. Aspect, P. Grangier, and G. Roger, Experimental tests of realistic local theories via Bell's theorem, Phys. Rev. Lett. **47**, 460 (1981).

[37] S. H. Strogatz, *Nonlinear Dynamics and Chaos: With Applications to Physics, Biology, Chemistry and Engineering* (Westview Press, Boulder, CO, USA, 2000).